\documentstyle[multicol,prb,aps,epsfig]{revtex}

\def\bigup{\tilde{\uparrow}}
\def\bigdown{\tilde{\downarrow}}
\def\SS{\tilde{S}}
\def\bigh{\tilde{h}}
\def\bigm{\tilde{m}}

\begin{document}

\title{Coupled Ladders in a Magnetic Field}

\author{T. Giamarchi}
\address{Laboratoire de Physique des Solides, CNRS URA 02,
U.P.S. B\^at 510, 91405 Orsay, France}
\author {A. M. Tsvelik}
\address{Department of Physics, University of Oxford, 1 Keble Road,
Oxford, OX1 3NP, United Kingdom}
\date{\today}

\maketitle

\begin{abstract}
We investigate the phase transitions in two-legs ladder systems
in the incommensurate phase, for
which the gap is destroyed by a magnetic field ($h_{c1}< h$) and the
ladder is not yet totally saturated ($h < h_{c2}$).
We compute quantitatively the
correlation functions as a function of the magnetic
field for an isolated strong coupling ladder $J_\perp \gg J_\parallel$
and use it to study
the phase transition occuring in a three
dimensional array of antiferromagnetically coupled ladders. The three
dimensional ordering is in the universality class of Bose condensation
of hard core bosons. We compute the critical temperature $T_c(h)$
as well as various physical quantities such as the NMR relaxations
rate. $T_c$ has an unusual camel-like shape with a local minimum at
$h=(h_{c1}+h_{c2})/2$ and behaves as $T_c \sim (h-h_{c1})^{2/3}$ for
$h\sim h_{c1}$. We discuss the experimental consequences for compounds
such as Cu$_2$(C$_5$H$_{12}$N$_2$)$_2$Cl$_4$
\end{abstract}

\pacs{PACS numbers:}
\begin{multicols}{2}

\section{Introduction}

There has been recently considerable interest
\cite{dagotto_2ch_review} on spin ladder materials.
These systems, quite remarkably, have a gap in the spin excitation
spectrum for an even number of legs and no gap for an odd number.
This phenomenon, reminiscent of the Haldane conjecture
\cite{haldane_gap,schulz_spins} has been
explored in great details both theoretically
\cite{strong_spinchains,gopalan_2ch,shelton_spin_ladders,white_2ch,%
sandvik_srcuo,hida_2ch,poilblanc_4ch,greven_2ch}
and experimentally \cite{takano_spingap,chiari_cuhpcl,chaboussant_cuhpcl,%
hammar_cuhpcl,carter_lacasrcuo}.

In ladders, contrarily to the case of spin $S$ chains,
the gap and the dispersion in the ladder
are controlled by two different energy scales, namely the transverse
$J_\perp$ and longitudinal $J_\parallel$ exchanges.
The ladders are thus prime candidates to
study quantum phase transitions where the spin gap is destroyed by
application of a magnetic field. Because of this separation of energy
scales between the gap and the exchange, even when the gap is
destroyed quantum effects are still crucial. The ladders thus offers
the possibility of an extremely rich quantum behavior, unsuspected
in more conventional spin systems.
Such quantum phase transitions were indeed studied experimentally
\cite{chaboussant_cuhpcl,chaboussant_nmr_ladder,%
chaboussant_ladder_strongcoupling}. On the theoretical side
they were investigated using a bosonization technique
\cite{chitra_spinchains_field}
for a single ladder. Close to the critical point where the gap vanished,
the spin-spin correlation functions were found to diverge with a
universal exponent, leading
to a divergent NMR relaxation rate $1/T_1 \sim T^{-1/2}$, in good
agreement with the experimental findings. Between the critical field
$h_{c1}$ where the gap was destroyed and the saturation field $h_{c2}$,
the ladder had incommensurate spin spin correlation function with a
quite distinctive spectrum compared to single chain systems.
These results were
confirmed and extended in subsequent analytical and numerical calculations
\cite{totsuka_ladder_strongcoupling,mila_ladder_strongcoupling,%
furusaki_correlations_ladder,usami_numerics_ladder}.

Due to the gaped nature of the excitations for a single ladder
when $h < h_{c1}$ a weak interladder coupling is
inefficient and the single ladder approximation is nearly
exact. This is clearly different in the incommensurate phase
$h_{c1} < h < h_{c2}$, and the question of the coupling of ladders
becomes much more crucial. Quite generally interladder
coupling can lead now to a three dimensional ordered phase.
This is the case for example for the compound
Cu$_2$(C$_5$H$_{12}$N$_2$)$_2$Cl$_4$ which has an
experimentally accessible gap of $\Delta \sim 11K$.
Specific heat measurements
have revealed the existence of a transition at finite temperature,
the nature of which is still controversial
\cite{chaboussant_ladder_strongcoupling,hammar_transition_cuhpcl,%
calemczuk_heat_ladder,nagaosa_lattice_ladder}.
It is thus a challenge, both from a theoretical point of view and in
view of application to experiments, to understand how three
dimensional ordering can occur in ladder systems.

We investigate the nature and physical properties of such transition
by looking at antiferromagnetically coupled ladders.
Because of the peculiar nature of the
excitation spectrum in ladders, this transition is different from
the one occuring in more conventional spin materials.

The plan of the paper is as follows. In section~\ref{sec:model} we
define the model for coupled ladders. For simplicity we confine
ourselves to the case of strongly coupled ladders $J_\perp \gg
J_\parallel$. In section~\ref{sec:single} we examine the single ladder
in this limit, using a mapping on a single spin chain
\cite{totsuka_ladder_strongcoupling,chaboussant_ladder_strongcoupling,mila_ladder_strongcoupling}.
We compute quantitatively the correlation functions as a function of
the magnetic field.
The weak and strong coupling limits give an identical structure for
the correlations functions and we recover the universal exponents and
spectrum for the spin-spin correlation functions derived in
Ref.~\onlinecite{chitra_spinchains_field}.
The three dimensionally coupled ladders are described in
section~\ref{sec:coupled}. The three dimensional ordering is in the
universality class of Bose condensation.
We compute the critical temperature $T_c$
as well as various physical quantities such as the NMR relaxations
rate. $T_c$ has a camel-like shape with a local minimum at
$h=(h_{c1}+h_{c2})$ and behaves as $T_c \sim (h-h_{c1})^{2/3}$ for
$h\sim h_{c1}$. We discuss the experimental consequences for compounds
such as Cu$_2$(C$_5$H$_{12}$N$_2$)$_2$Cl$_4$.
Conclusions can be found in section~\ref{sec:conclusion} and some
technical details are left for the Appendix.

\section{The model}  \label{sec:model}

\begin{figure}\narrowtext
\centerline{\epsfig{file=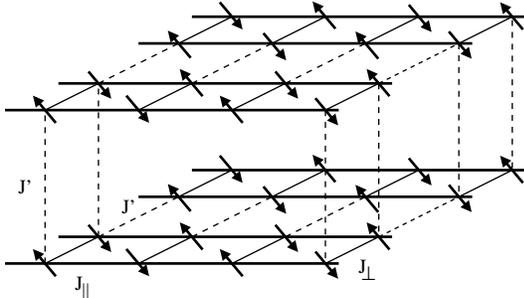,angle=-90,width=7cm}}
\caption{\label{fig:ladder}
The two leg ladder system considered in this paper.
An interladder coupling $J'$ couples the ladder in a
three dimensional way.}
\end{figure}

We consider the two legs ladders shown in Figure~\ref{fig:ladder}.
For the moment we consider a single ladder and thus take $J'=0$.
The ladder Hamiltonian is given by
\begin{equation} \label{eq:ladderham}
H = J_\parallel \sum_{i,l=1,2} \vec{S_{i,l}} \cdot \vec{S_{i,l}}
 + J_\perp \sum_i \vec{S_{i,1}} \cdot \vec{S_{i,2}} - h \sum_{i,l=1,2}
S^z_{i,l}
\end{equation}
where $l=1,2$ denote the two legs of the ladder, and $h$ the applied
magnetic field.

The case when the rung  coupling $J_{\perp}$ is much
smaller than the interaction $J_{\parallel}$ along the ladder has
been studied by a variety of techniques both in the absence of
\cite{schulz_spins,strong_spinchains,gopalan_2ch,shelton_spin_ladders,%
white_2ch,sandvik_srcuo,hida_2ch,greven_2ch}
or in the presence of a magnetic field
\cite{chitra_spinchains_field,furusaki_correlations_ladder}
We concentrate here on the
opposite limit $J_\perp \gg J_{\parallel}$. In that case
the ladder can be mapped onto a single spin 1/2 chain
\cite{totsuka_ladder_strongcoupling,chaboussant_ladder_strongcoupling,mila_ladder_strongcoupling},
and we recall the mapping here for completeness.
Indeed an  individual rung may be in a  singlet
or a triplet state. Applying a magnetic field brings one component of
the triplet closer to the singlet ground state such that for a strong
enough magnetic field we have a situation when singlet and $m = -1$
component of triplet create a new effective spin-1/2. It is thus
possible if $J_\perp \gg J_\parallel$ to retain only these two states
for all the magnetic field range between $h_{c1}$ when the gap is broken
to $h_{c2}$ when the ladder is completely magnetized.

One can easily project the original Heisenberg Hamiltonian (\ref{eq:ladderham})
on the new singlet-triplet subspace
\begin{eqnarray}
|\bigdown\rangle & = &
   \frac1{\sqrt2}[|\uparrow\downarrow\rangle - |\downarrow\uparrow\rangle] \\
|\bigup\rangle & = & |\uparrow\uparrow\rangle    \nonumber
\end{eqnarray}
This leads to the definition of the effective spin 1/2 operators
\begin{eqnarray} \label{eq:effective}
S^+_{1,2} &=& \mp \frac1{\sqrt2} \SS^+ \\
S^z_{1,2} &=& \frac14 [I + 2 \SS^z]
\end{eqnarray}

When expressed in term of the effective spin operators (\ref{eq:effective}),
the original Hamiltonian (\ref{eq:ladderham}) becomes
\begin{eqnarray}
H_{\text{eff}} & = &
J_\parallel \sum_i [\SS^x_i \SS^x_{i+1} + \SS^y_i \SS^y_{i+1} +
\frac12  \SS^z_i \SS^z_{i+1}] \nonumber \\
 & & - \bigh \sum_i \SS^z_i + C \label{eq:hameffective}
\end{eqnarray}
where $C = (-\frac{J_\perp}4 + \frac{J_\parallel}8 - \frac{h}2) L$
is a simple energy shift and the system is in an effective magnetic
field
\begin{equation}
\bigh = h - J_\perp - \frac{J_\parallel}2
\end{equation}
The Hamiltonian
(\ref{eq:hameffective}) describes a single spin 1/2 chain with a \emph{fixed}
XY anisotropy of $1/2$ in an effective magnetic field.
In the following we denote with a tilde, the magnetic field $\bigh$, and
the magnetization $\bigm$ of the effective spin 1/2 chain. The gaped
phase $h < h_{c1}$ for the ladder corresponds to the
negatively saturated magnetized phase for the effective spin chain, whereas the
massless phase for the ladder corresponds to the finite magnetization phase
for the effective spin 1/2 chain
\cite{chaboussant_ladder_strongcoupling}.
The field $h_{c2}$ where the ladder is
totally magnetized correspond to the fully magnetized phase for the
effective spin 1/2 chain. It is easy to check that
\begin{equation}
\bigh_{c1,c2} = \mp \frac{3 J_\parallel}2
\end{equation}

\section{Single ladder} \label{sec:single}

Before taking into account interladder interactions let us first recall some
important consequences of such a mapping for the single ladder. In the
process we give a more quantitative calculation for the correlation
functions as a function of the magnetic field. The results of this
section will be used to study the interladder coupling.
We focus here on the massless phase $h_{c1} < h < h_{c2}$.
To conveniently derive the low energy properties of the effective spin
1/2 chain we use the by now well known bosonization technique.
We refer the reader to
Ref.~\onlinecite{emery_revue_1d,solyom_revue_1d,affleck_houches,%
chitra_spinchains_field}
for details and just recall here the main steps.

\begin{figure}
\centerline{\epsfig{file=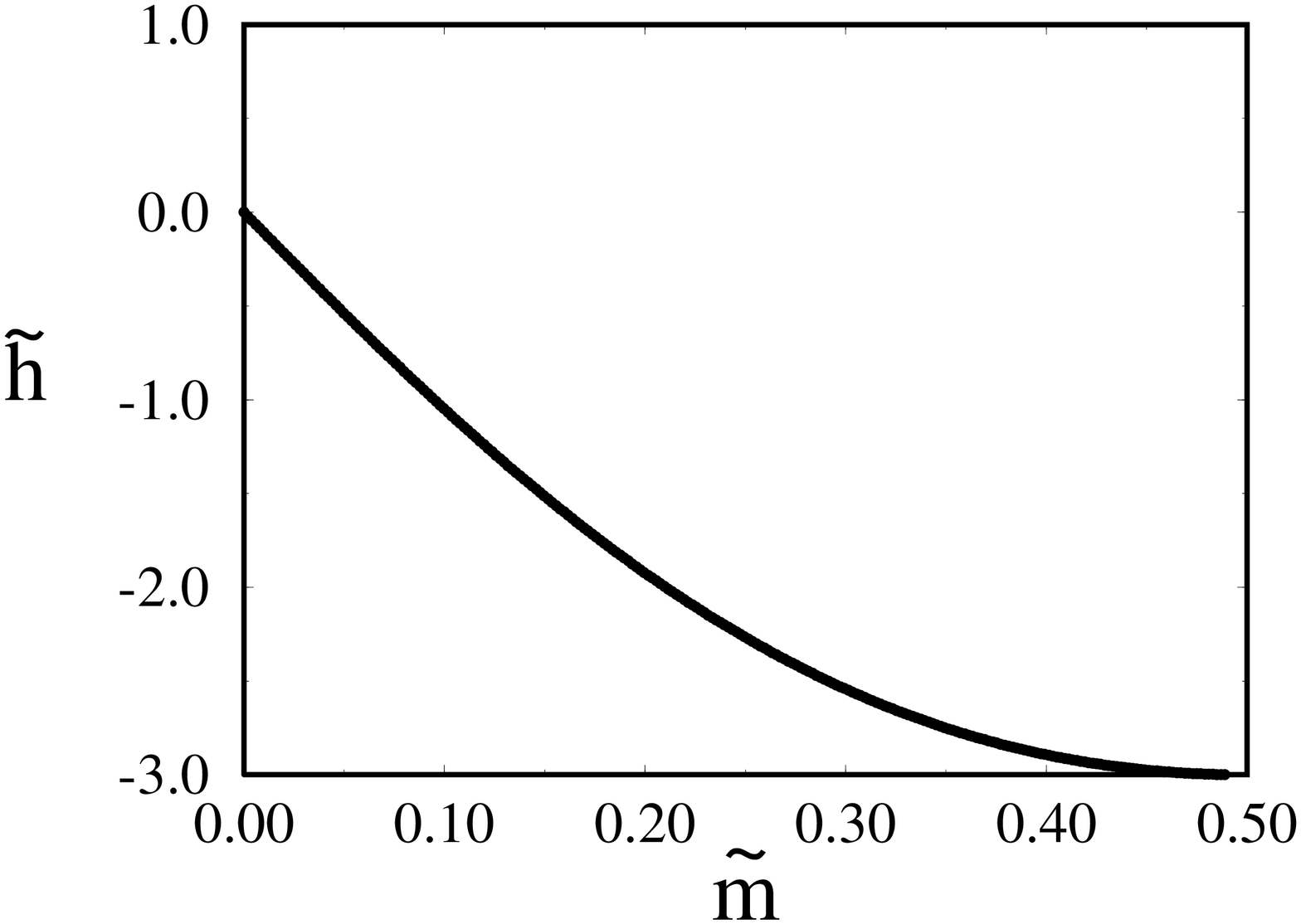,angle=-90,width=7cm}}
\centerline{\epsfig{file=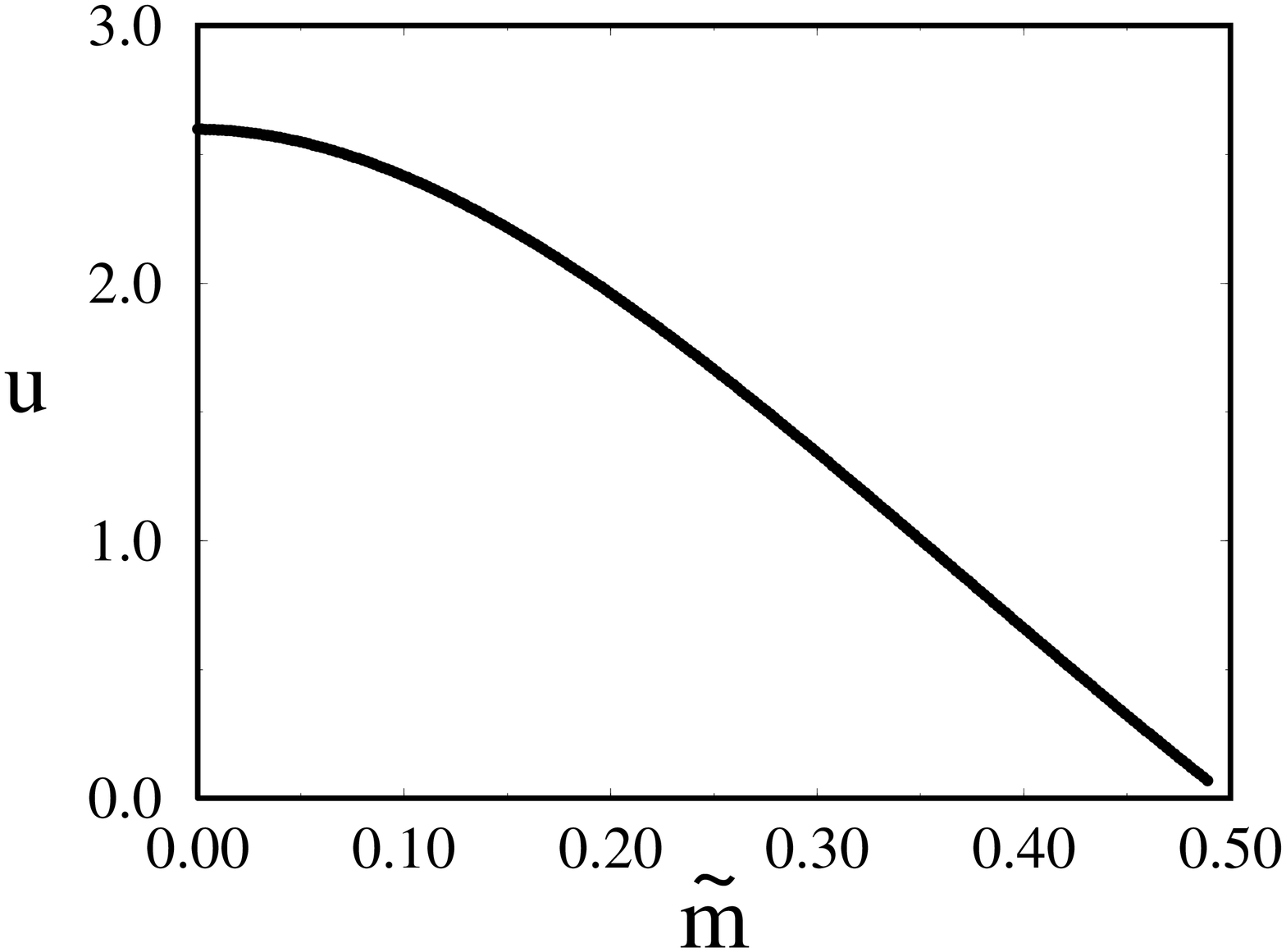,angle=-90,width=7cm}}
\centerline{\epsfig{file=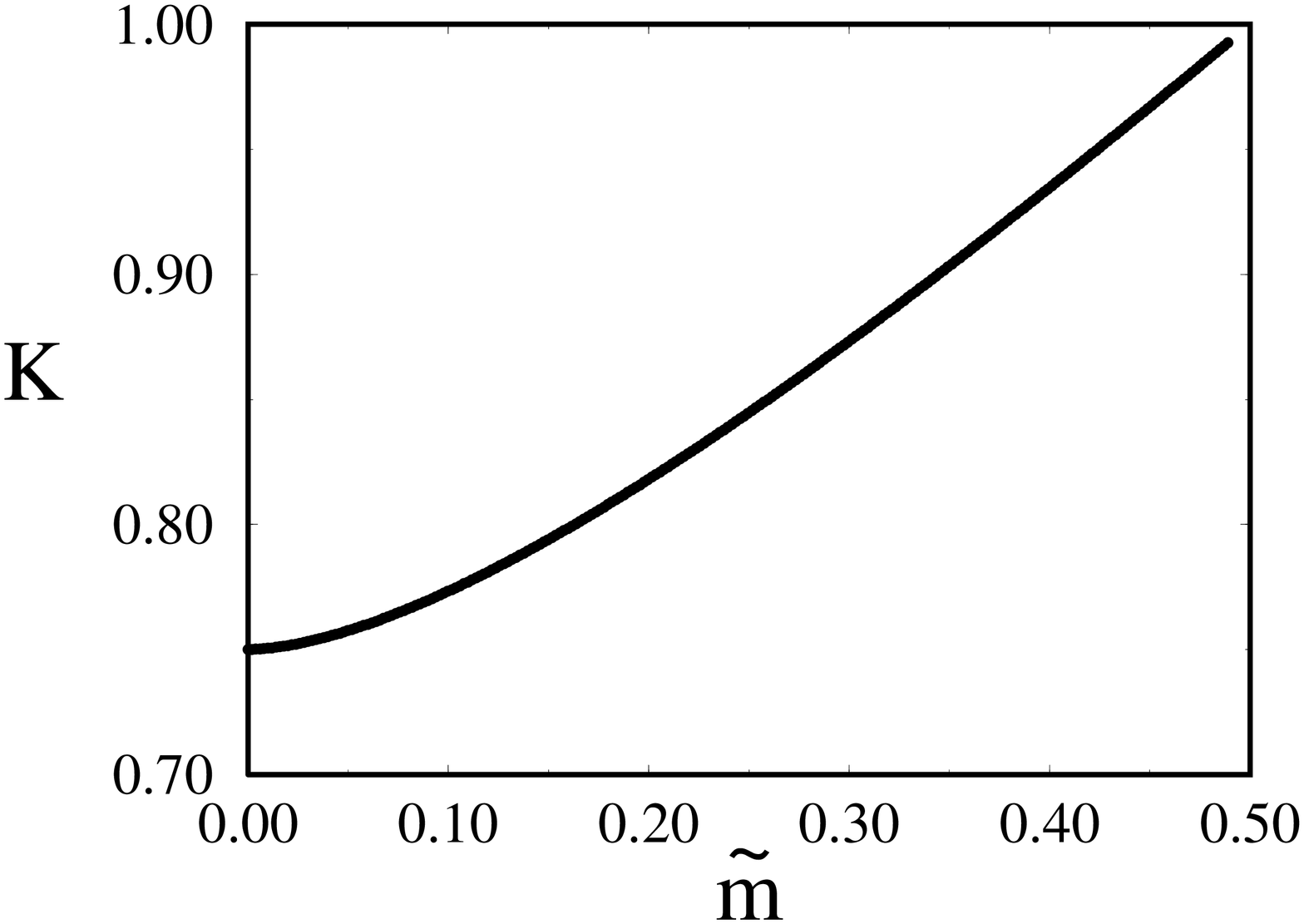,angle=-90,width=7cm}}
\caption{\label{fig:lutparam}
Magnetic field $\bigh$ and Luttinger liquid parameters $u$ and $K$
for an XY anisotropy of $1/2$ plotted as a function of the
magnetization $\bigm$ (only positive values are shown, the parameters
being symmetric with $\bigm \to -\bigm$).
$\bigm=1/2$ is the saturated chain. $K=3/4$ for zero
magnetic
field, whereas $K\to 1$  and $u \to 0$ close to saturation since
the excitations above the ground state become very diluted.
$u$ and $\bigh$ are in units of $J_\parallel/2$.}
\end{figure}

We first use the
Jordan-Wigner transformation
\cite{affleck_houches,jordan_transformation,nijs_equivalence} which
essentially  maps the
spin problem onto a problem of interacting fermions on a lattice.
For the spin 1/2 system considered here, the corresponding
fermionic problem has Fermi momentum $k_F = {\pi \over 2}$ if $\bigh=0$.
Finite magnetic field corresponds to a chemical potential for the
fermions. We then perform a linearization  around the free Fermi points
given by $\pm k_F$ , to obtain  an  effective low energy continuum
fermionic theory and then express the fermion operators in term of
bosonic ones related to the fermion density fluctuations
using the standard dictionary of Abelian bosonization.
\begin{eqnarray} \label{eq:bosonizedspin}
S^{+}(x) & = &\frac{e^{-\imath \theta(x)}}{\sqrt{2\pi a}}
\left[e^{-\imath \frac{\pi x} a}+\cos 2\phi(x) \right]\nonumber \\
S_z(x) & = & -\frac{1}{\pi}\partial_{x}\phi +
e^{\imath \frac{\pi x} a} \frac{ \cos 2\phi(x)}{\pi a}
\end{eqnarray}
Where $S^{+}(x)=\frac{S^{+}_n}{\sqrt{a}}$, $S^z(x)=\frac{S^z_n}{a}$ for
$x=na$, $a$ being the distance between two nearest neighbors sites
along the chain. From now on we take the lattice spacing $a=1$ and
measure all distance in units of $a$.
The phase $\phi$ is related to the average density of fermions (or
equivalently to the uniform spin density along $z$)
by $S_z(x) = -\frac{1}{\pi}\partial_{x}\phi$, whereas $\theta$ is
connected to the conjugate momentum $\Pi$ of $\phi$ (such that
$[\phi(x),\Pi(x')] = i \delta(x-x')$) by $\theta(x) = \int_{-\infty}^x
dy \Pi(y)$. In a very crude sense $\phi,\theta$ can be viewed as the
polar angles of a spin. The low energy properties
of the Hamiltonian (\ref{eq:hameffective}) can be totally described in
terms of the boson Hamiltonian
\begin{equation} \label{eq:hbosons}
H = \int \frac{dx}{2\pi}\left[ u K (\pi \Pi)^2
+\frac{u}{K}(\partial_{x}\phi)^2
\right]
\end{equation}
where $\phi$ has been shifted to absorb the finite magnetization
\begin{equation}
\phi \to \phi + 2 \bigm x
\end{equation}
The only two parameters controlling the low energy properties are the
``spin wave'' velocity $u$ and a number $K$ called the Luttinger liquid
exponent. Both are known exactly for the spin 1/2 chain
\cite{haldane_xxzchain}. For $\bigh=0$ analytic expressions are known
\begin{eqnarray}
J_z/J_{x,y} &=& - \cos\pi\beta^2 \nonumber \\
1/K &=& 2 \beta^2  \label{eq:lutcalc} \\
u & =& \frac1{1-\beta^2} \sin(\pi(1-\beta^2))\frac{J_\parallel}2 \nonumber
\end{eqnarray}
Thus  $K=1/2$ for an isotropic
Heisenberg chain with $\bigh=0$  whereas $K=1$ for the pure XY one.
For the Hamiltonian (\ref{eq:hameffective}) this leads to
\begin{equation}
K = 3/4 \qquad,\qquad u=\frac{3\sqrt3}2 \frac{J_\parallel}2
\end{equation}
At finite magnetic field $K$ and $u$ can be obtained by integration of
the Bethe ansatz equations  and are shown on
Figure~\ref{fig:lutparam} for the specific case of the XY anisotropy
$1/2$.
Close to $h_{c1}$ or $h_{c2}$ the number of excitations compared to the
fully polarized ground state becomes very small (in the fermionic
language one is close to an empty or a full band), and thus $K$ take the
value for noninteracting particles $K \to 1$ regardless of the strength
of the original interaction $J_z/J_{xy}$.

\begin{figure}
\centerline{\epsfig{file=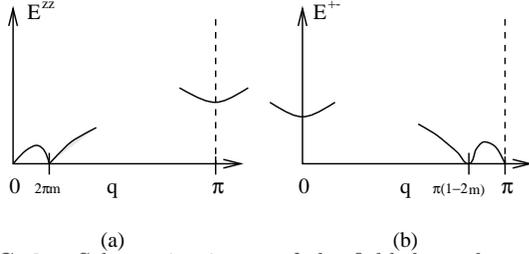,angle=-90,width=7cm}}
\caption{\label{fig:masslessmassive} \narrowtext
Schematic picture of the field dependent dispersion as seen by the
$S^zS^z$ (a) and $S^+S^-$ (b) correlations for fields close to $h_{c1}$.
Only the dominant singularities are shown.
In marked contrast with a single chain there is no massless excitations
close to $q \sim \pi$ (resp. $q \sim 0$) due to the presence of an
antisymmetric massive mode, unaffected by the magnetic field
\protect{\cite{chitra_spinchains_field}}.}
\end{figure}

Since the free boson theory given by (\ref{eq:hbosons}) is trivially
solvable, it
is straightforward to calculate the asymptotic decay of the
dynamic correlation functions, which are just the ones of a spin 1/2 chain.
Using (\ref{eq:bosonizedspin}), one gets for $T=0$
(for more details see e.g. Ref.~\onlinecite{chitra_spinchains_field})
\end{multicols} \widetext
\begin{eqnarray}
\langle \SS^z (x,\tau) \SS^z (0,0) \rangle & =
 & {\bigm}^2 + C_1 \frac1{r^2} + C_2 \cos(\pi(1-2\bigm)x)
\left(\frac1r\right)^{2K}
\nonumber \\
\langle \SS^+ (x,t) \SS^- (0,0) \rangle & =  &  C_3 \cos (2\pi \bigm x)
\left(\frac1r\right)^{2K + 1/(2K)}
+ C_4 \cos (\pi x) \left(\frac1r\right)^{1/(2K)}
\end{eqnarray}
where $r = \sqrt{x^2 + (u \tau)^2}$ and $C_i$ are constants on
which we focus later in this section.
When expressed in term of the \emph{true} magnetization
$2 m = 1 + 2 \bigm$ and the original spin operators of the ladder
using (\ref{eq:effective}) this gives (e.g. for rung $1$)
\begin{eqnarray} \label{eq:corladderstrong}
\langle S^z_1 (x,t) S^z_1 (0,0) \rangle & =
 & \frac{m^2}4 + \frac1{r^2} + \cos(2 \pi m x) \left(\frac1r\right)^{2K}
\nonumber \\
\langle S^+_1 (x,t) S^-_1 (0,0) \rangle & =  &  \cos (\pi(1-2m) x)
\left(\frac1r\right)^{2K + 1/(2K)}
+ \cos (\pi x) \left(\frac1r\right)^{1/(2K)}
\end{eqnarray}
\begin{multicols}{2}
(where we have dropped the constants $C$ for simplicity).
Equ. (\ref{eq:corladderstrong}) presents some remarkable features.
First, as already pointed out in Ref.~\onlinecite{chitra_spinchains_field},
low energy modes appear only close to $q\sim 0$ in the $S^z$
correlation function or close to $q\sim \pi$ for the transverse one.
The $q\sim \pi$ (for $S^z$) or $q \sim 0$ (for $S^\pm$) mode remain
\emph{massive}. This is in marked contrast to what would happen for a
gaped (e.g. dimerized or frustrated) single chain system
(in weak coupling) where
\emph{both} the $q \sim 0$ and $q\sim \pi$
would become massless when $h \geq h_{c1}$. A summary of the massless
and massive modes is shown on Figure~\ref{fig:masslessmassive}
Such prediction for the ladder correlations should be
testable in neutron or resonance experiments.
Close to $h_{c1}$ (or $h_{c2}$) $K \to 1$ and one recovers
the universal exponent for the decay of the correlation functions predicted
in Ref.~\onlinecite{chitra_spinchains_field}. The weak coupling approximation
only allowed for a qualitative
calculation of the exponents far from $h_{c1}$ and $h_{c2}$.
For the strong coupling case one
can get the full magnetic field dependence as shown in
Figure~\ref{fig:lutparam}. As
shown in Appendix~\ref{ap:weaktostrong}, the \emph{whole asymptotic}
structure of the correlation function is independent of the strength
of the coupling $J_\perp$ vs $J_\parallel$
provided of course that the correct Luttinger liquid
exponent are used.

\begin{figure}
\centerline{\epsfig{file=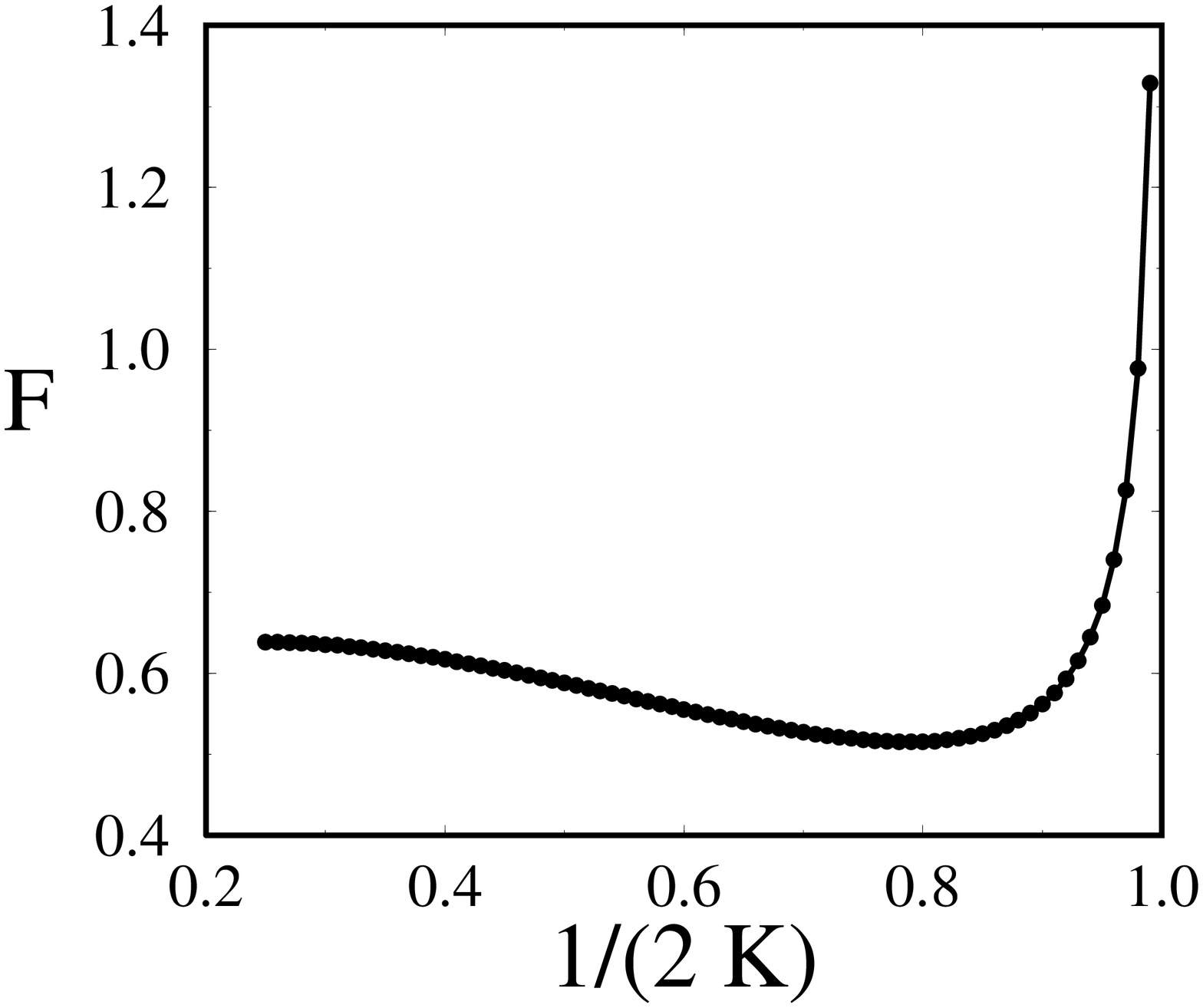,angle=-90,width=7cm}}
\centerline{\epsfig{file=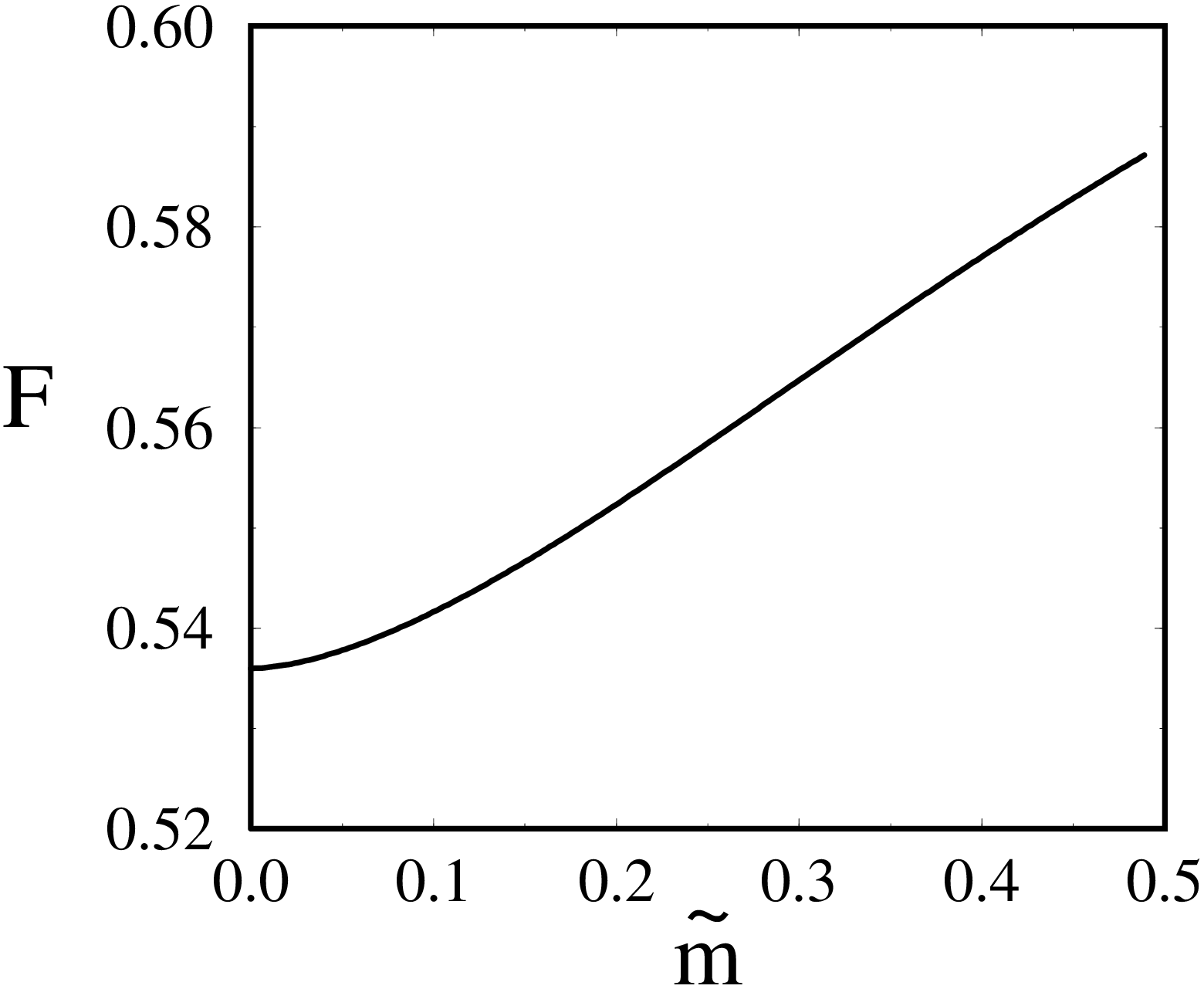,angle=-90,width=7cm}}
\caption{\label{fig:prefplot}
Prefactor $F$ as a function of the Luttinger
parameter $K$. The divergence close to $K=1/2$ is due
to the appearance of logarithmic corrections in the correlations
functions for the isotropic case \protect{\cite{giamarchi_logs}}.
For the specific case of the chain with XY anisotropy $1/2$, the
prefactor is shown as a function of the magnetization $m$. Note
the relatively weak dependence on the whole interval due to
the finite XY anisotropy of the effective spin chain.}
\end{figure}

Let us now compute quantitatively the correlation functions.
We only focus here on the massless modes.
The prefactors in the correlation functions of the spin
operators in the XXZ  model have been
computed by Lukyanov and Zamolodchikov
\cite{lukyanov_sinegordon_correlations,lukyanov_spinchain_correlations}.
Thus for example for the transverse staggered magnetization we have
\begin{equation} \label{eq:spinq}
(-1)^n\SS^+(x = na,\tau) =
[F_{\beta}/8]^{1/2}(a/u)^{\beta^2/2} e^{i\theta(x,\tau)}
\end{equation}
where the expression for the prefactor
reads \cite{lukyanov_spinchain_correlations}
\begin{eqnarray}
F &=& \frac{1}{2(1 - \beta^2)^2}\left[\frac{\Gamma(
\frac{\beta^2}{2 - 2\beta^2})}
{2\sqrt\pi \Gamma(\frac{1}{2 - 2\beta^2})}\right]^{\beta^2}\times
\label{pref} \\
& & \exp\left\{- \int_0^{\infty}\frac{d t}{t}\left(\frac{\sinh(\beta^2t)}
{\sinh t\cosh[(1 - \beta^2)t] } - \beta^2 e^{-2t}\right)\right\}
\nonumber
\end{eqnarray}
In the vicinity of the value of interest here $\beta^2 = 2/3$ an
analytic calculation is possible:
\begin{eqnarray}
F_{\beta} &=&
\frac{9}{2\pi^{2/3}}\left[\frac{\Gamma(2/3)}{\Gamma(1/3)}\right]^2
e^{ - (\beta^2 - 2/3)[\gamma + \ln (3\pi/16) + \pi/\sqrt 3]}
\nonumber \\
&\approx&  F(2/3) \exp[ - 2.173(\beta^2 - 2/3)]
\end{eqnarray}
For more general values of $K$ and $m$ the value of $F$ is shown on
 Figure~\ref{fig:prefplot}
Using (\ref{eq:spinq}) one gets for the
staggered part of the transverse spin-spin correlation
function of the physical ladder spins  as a function of
space, time and temperature
\begin{equation} \label{eq:decay}
\langle S^+(na, \tau)S^-(0,0) \rangle =
\frac{F_{\beta}}{16}\left(\frac{a}{u}\right)^{\beta^2}\left|
\frac{\pi T}{\sin\pi T(\tau + i x/v)}\right|^{\beta^2}(-1)^Q
\end{equation}
The term $(-1)^Q$ indicates that due to  the relation between physical
and effective
spins (\ref{eq:effective}),
the physical correlation function is singular at the wave vector $(\pi,\pi)$.
Let us insist that the position of the  singularity does not
change with field \cite{chitra_spinchains_field}
(see (\ref{eq:corladderstrong})).

Assuming that the local susceptibility is dominated by the
contribution from the
transverse part of the staggered susceptibility, we get the following
expression for the NMR relaxation rate:
\begin{equation} \label{eq:nmrbooks}
\frac1{T_1} \propto
T\lim_{\omega \to 0}\frac{\chi_{\text{loc}}''(\omega)}{\omega}
\end{equation}
(up to the hyperfine coupling constants).
This leads, when the Fourier transform and analytical continuation
of (\ref{eq:decay}) is performed (see also (\ref{eq:suc1d})), to the
relaxation rate
\begin{equation} \label{eq:nmrbase}
\frac1{T_1} \propto
\frac{F_{\beta}}{8}
\Gamma^2(\beta^2/2)
\Gamma(1 - \beta^2)\frac{T}{(2\pi T)^{2 - \beta^2}}
\end{equation}
At $\beta^2 = 2/3$ (i.e. for $\bigh =0$ or $h = (h_{c1}+h_{c2})/2$)
this gives for  $T_1$
\begin{eqnarray}
\frac1{T_1} & \propto &
T \lim_{\omega \to 0}\frac{\chi_{loc}''(\omega)}{\omega} \\
 & = &
\frac{1}{T^{1/3}}\left(\frac{u}{a}\right)^{1/3}
\frac{3\sqrt 3\Gamma(2/3)}{16\sqrt 2\pi}
\approx 0.1\frac{1}{T^{1/3}}\left(\frac{v}{a}\right)^{1/3}
\end{eqnarray}
Close to $h_{c1}$ or $h_{c2}$, $K\to 1$ (thus $\beta^2 \to 1/2$) and
(\ref{eq:nmrbase}) gives back the universal
\cite{chitra_spinchains_field} divergence of the
relaxation time
\begin{equation}
\frac1{T_1} \propto \frac1{T^{1/2}}
\end{equation}
Away from
the critical field the exponent increases weakly to $-1/3$ at $\bigh=0$.
(see also
Ref~\onlinecite{mila_ladder_strongcoupling,usami_numerics_ladder}).
The full magnetic field dependence can be obtained from
Figure~\ref{fig:lutparam}
(using $\beta^2 = 1/2K$ and (\ref{eq:nmrbase})).

\begin{figure}
\centerline{\epsfig{file=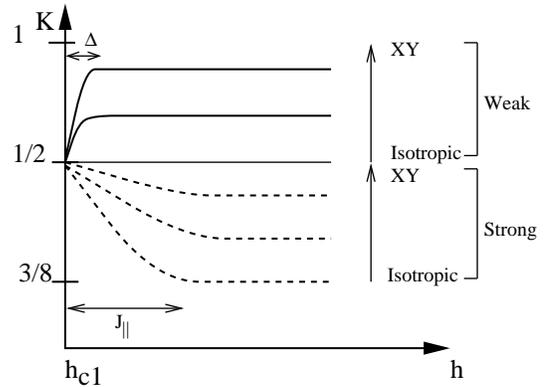,angle=-90,width=7cm}}
\caption{\label{fig:weaktostrong}
Qualitative variation of the Luttinger liquid parameter $K$ as a
function of the magnetic field both for the weak and strong coupling
ladder. Full line is the weak coupling ladder $J_\perp \ll
J_\parallel$, and dashed line for the strong coupling one $J_\perp \gg
J_\parallel$. Different curves correspond to different XY anisotropy
from the isotropic case to the XY limit. $\Delta$ is the gap for the
weak coupling case. See also Figure~\ref{fig:lutparam} for the exact result
for the strong coupling ladder (due to the difference
of definition (see text) between $K_{\text{weak}}$ and
$K_{\text{strong}}$ we plot $K_{\text{strong}}/2$).}
\end{figure}

Although, as shown in Appendix~\ref{ap:weaktostrong}, the
correlation functions for the strong and weak coupling ladder are
smoothly connected, a very interesting question is how the Luttinger
liquid parameter varies with the field when going from weak to strong
coupling. This is not trivial since for the weak coupling ladder
when $\Delta \ll h \ll J_\parallel$ one recovers essentially the
Luttinger liquid exponent of a single chain\cite{chitra_spinchains_field}.
For an isotropic
system this is $K=1/2$, i.e. the same value than the universal one
close to $h_{c1}$. If there is XY anisotropy the parameter $K$
increases and thus  $K \geq 1/2$.
On the other hand for the strongly coupled ladder \emph{regardless} of
the XY anisotropy (provided $J_z>0$) the luttinger liquid parameter
decreases with the field (see (\ref{eq:hameffective},\ref{eq:lutcalc})
and Figure~\ref{fig:lutparam}), giving the very different field
dependence shown in Figure~\ref{fig:weaktostrong}.
It would be extremely interesting to have  more quantitative estimates
for the behavior shown in
Figure~\ref{fig:weaktostrong}. An amusing possibility would be to have,
for a certain strength of coupling and a given anisotropy a luttinger
liquid parameter totally field independent.

\section{Coupled ladders} \label{sec:coupled}

In order to describe realistic compounds we now take into account an
interladder interaction of the form shown in Figure~\ref{fig:ladder}
and given by the Hamiltonian
\begin{equation}
H_{3D} = \sum_\alpha H_{\text{ladder}}^\alpha +
J' \sum_{\langle\alpha,\beta\rangle}
\sum_i \vec{S_{i,\alpha,1}} \cdot \vec{S_{i,\beta,2}}
\end{equation}
where $\langle \alpha,\beta \rangle$ denotes pairs of nearest neighbors
ladders.
It is easy to see from Figure~\ref{fig:ladder} that a spin on leg $1$
of one ladder can only interact with the spin on leg $2$ of the neighboring
ladder and vice versa.

Since the interladder coupling is very weak it is again legitimate
to map the problem to an effective spin 1/2 problem. The coupled ladder
system thus reduces to a problem of spin 1/2 chains coupled
by the interaction
\end{multicols} \widetext
\begin{equation} \label{eq:3dcoupling}
H_{\text{coupling}} = -\frac{J'}4 \sum_{\langle\alpha,\beta\rangle}
                      [\SS^+_\alpha \SS^-_\beta + \text{h.c.}]
                     +\frac{J'}4 \sum_{\langle\alpha,\beta\rangle}
                     \SS^z_\alpha \SS^z_\beta
                     +\frac{J'z}8 \sum_\alpha \SS^z_\alpha
\end{equation}
\begin{multicols}{2}
where $z$ is the coordination number.
Because a spin on leg $1$ can only be coupled to a spin on leg $2$ by
$J'$ this leads to a \emph{ferromagnetic} coupling for the XY part of the
interchain coupling although the original interladder coupling is
antiferromagnetic. There is also a trivial redefinition of the
effective magnetic field by the interladder coupling.
Although the problem of coupled ladders now look identical
to the one of three dimensionally coupled spin 1/2 chain,
the physics will be quite different from the standard case
\cite{schulz_coupled_spinchains}
of \emph{isotropic} coupled spin 1/2 chains. Indeed, as we
will see below, the XY anisotropy of the effective spin 1/2
chain inherent in the ladder problem, plays a crucial
role.
The treatment of (\ref{eq:3dcoupling}) depends crucially on
what is the characteristic energy scale of the interladder
coupling when compared to what happens for a single ladder.

\subsection{High density limit}

\begin{figure} \narrowtext
\centerline{\epsfig{file=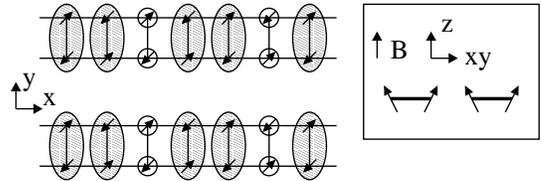,angle=-90,width=7cm}}
\caption{\label{fig:pictorial}
Cartoon of the three dimensional transition in ladder systems. The
direction of the spins in the XY plane tends to lock together between
different chains leading to a planar antiferromagnet. Note that
the triplet states
are in fact delocalized on the ladder and that in the XY plane
the spins remains modulated at $q=\pi$ in the ladder direction.
We have represented the singlets as shaded boxes.
This transition is in the universality class of normal-superfluid
transition or Bose-condensation of hard core bosons.}
\end{figure}

If one is far enough from $h_{c1}$ and $h_{c2}$, the interladder
coupling will be small compared
to the characteristic energies (Fermi energy for the associated spinless
fermion problem) of the single chain. It is then possible to treat the
interladder coupling in the mean field approximation while keeping the
full single ladder physics. Since the
single chain correlation functions along $z$ decays
faster than the one in the
XY plane (see e.g. (\ref{eq:corladderstrong}), three dimensional
order will occur \emph{first} in the XY plane.
It is thus possible to neglect in (\ref{eq:3dcoupling}) the interchain
$\SS^z \SS^z$ coupling, and to retain only the XY part.
Note that in that case it is not important whether the interchain
coupling
is ferro or antiferro since one can go from one to the other by making
the gauge transformation $\SS^x \to -\SS^x$, $\SS^y \to -\SS^y$,
$\SS^z \to \SS^z$
on alternating chains. The coupling is just a spin flip term for the
spin which in a bosonic representation for the chain is just a hoping
term for the bosons (see below). Another way of viewing it is as a
Josephson coupling between the phases $\theta$ of the spins on
different chains. This shows that the transition is a
normal-superfluid type transition or alternatively is a
Bose-condensation transition for the hard core bosons associated with
the effective spins. In a pictorial level this says that the
orientation of the spins of the ladder in the XY plane want to lock in
the same direction as shown in Figure~\ref{fig:pictorial}.
Since we are in $d=3$ (i.e. above the critical dimension for the
quantum transition \cite{fisher_bosons_scaling}) the exponents are the
mean-field ones ($\nu=1/2$ and $\zeta=1$).

To compute the transition one can use a standard mean-field approach.
The transition temperature is given by
\begin{equation} \label{eq:meaneq}
\frac1{J'} = \chi_\perp(q=0,\omega=0,T)
\end{equation}
where $\chi_\perp$ is the single chain transverse staggered susceptibility
(wavevector $q$ is to be counted relative to $\pi$).
For the Luttinger liquid the susceptibility can be computed
\cite{cross_spinpeierls,schulz_correlations_1d}
\end{multicols} \widetext
\begin{eqnarray} \label{eq:suc1d}
\chi_{\perp}(\omega, q) &=& - \frac{F_{\beta}}{8}\frac{a}{v}\left
[\sin(\frac{\pi\beta^2}2)\left(\frac{2\pi aT}{u}\right)^{-2 + \beta^2}
B(\frac{\beta^2}4 - is_+,1 - \frac{\beta^2}2)
B(\frac{\beta^2}4 - is_-,1 - \frac{\beta^2}2) \right. \nonumber \\
& & \left. - \frac{\pi}{1 - \beta^2/2}\right]
\end{eqnarray}
\begin{multicols}{2}
where
\begin{equation}
s_{\pm} = \frac{\omega - vq}{4\pi T}
\end{equation}
and $B(x,y) = \Gamma(x)\Gamma(y)/\Gamma(x + y)$.

Solving (\ref{eq:meaneq}) with (\ref{eq:suc1d}) gives
the critical temperature
\begin{equation} \label{eq:thetc}
T_c = u\left(\frac{J'D_{\beta}}{16\pi v}\right)^{1/(2 - \beta^2)}
\end{equation}
with
\begin{equation}
D_{\beta} = F_{\beta}\sin(\pi\beta^2/2)\left[\frac{\Gamma(\beta^2/4)\Gamma(1 -
\beta^2/2)}{\Gamma(1 - \beta^2/4)}\right]^2
\end{equation}

\begin{figure} \narrowtext
\centerline{\epsfig{file=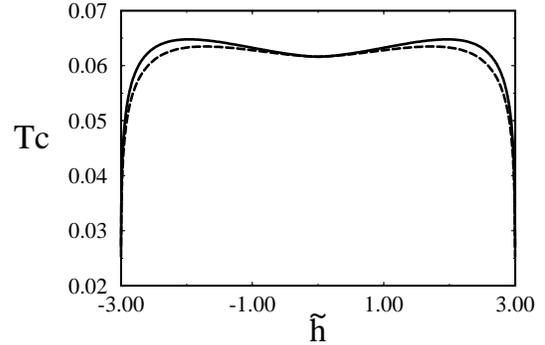,angle=-90,width=7cm}}
\caption{\label{fig:tclut}
Three dimensional transition temperature $T_c$ as a function of the
applied magnetic field $\bigh$ (i.e. for $h_{c1} < h < h_{c2}$).
The full line is the complete solution whereas
for the dotted curve the prefactor $F_\beta$ has been fixed to its
zero field value. Note the minimum at $\bigh = 0$}
\end{figure}

Given the fact that we used quantitative estimates for the
one-dimensional correlation functions and not just asymptotic
estimates (\ref{eq:thetc}) should even be able to give semi-quantitatively
the $T_c$ if $J'$ can be determined by an independent method.
Much more important however, is the field dependence of the $T_c$.
Indeed since the exponents depend on the field in a non trivial
way one can expect a non trivial magnetic field dependence of the
three dimensional transition temperature.
Close to $\bigh=0$ an analytical solution can be obtained.
We have
\begin{equation}
\frac{d\ln T_c/J'}{d H} = - \frac{3}{4}\frac{d\beta^2}{d H}
[1.16 + \ln(v/T_c a)] + \frac{1}{4}\frac{d \ln v}{d H}
\end{equation}
It is clear that at small enough $J'$ this expression becomes positive
and thus magnetic field increases the transition temperature.
A numerical estimate of $T_c$ is given in Figure~\ref{fig:tclut}.
The camel-type structure, instead of the naively expected dromedary one,
comes from the competition between the decrease
of the exponent when one moves closer to $h_{c1}$ or $h_{c2}$ (which
leads to an increase of the $T_c$), with the fact that the
excitation velocity $u$ decreases at the same time.

We can also obtain the local susceptibility close to $T_c$.
In the mean field approximation it is given by
\end{multicols} \widetext
\begin{equation}
\chi_{loc}(\omega) = \frac{1}{(2\pi)^2}\int
\frac{d q_1\d q_2}{[\chi_0^{-1} - 2J_1' - 2J_2']+ 4J_1'\sin^2q_1/2 +
4J_2'\sin^2q_2/2}
\end{equation}
When $\tau \equiv \chi_0^{-1}(0,0) - 2J_1' - 2J_2' \ll J'$ one
can linearize the sinuses and calculate the integral
\begin{equation}
\Im m\chi_{loc}(\omega) = \frac{1}{4\pi\sqrt{J_1'J_2'}}\tan^{-1}\left[
\frac{\Im m\chi_0^{-1}}{\Re e\chi_0^{-1} - 2J_1' - 2J_2'}\right]
\end{equation}
In the limit of zero frequency we get
\begin{equation} \label{eq:localsusc}
\lim_{\omega \rightarrow 0}\frac{\chi_{loc}''(\omega)}{\omega} =
\frac{1}{\pi u}\int
\frac{d x[\psi(1 - \beta^2/4 + ix) - \psi(\beta^2/4 + i x)]}{1 -
(T_c/T)^{2 - \beta^2}
\left|\frac{\Gamma(\beta^2/4 - ix)\Gamma(1 - \beta^2/4)}{\Gamma(1 -
\beta^2/4 - ix)
\Gamma(\beta^2/4)}
\right|^2}
\end{equation}
\begin{multicols}{2}
Close to $T_c$ (\ref{eq:localsusc}) reduces to a mean field divergence
$1/(T-T_c)^{1/2}$. This leads to a similar divergence in the
relaxation rate $1/T_1$.

\subsection{Low density limit}

When the magnetic field is close to $h_{c1}$ or $h_{c2}$ the above
mean field approach on the single chain Luttinger liquid cannot be
used. Indeed the energy of the interchain coupling becomes larger than
the intra chain energy scale, and interchain coupling should be treated from
the start. Fortunately the problem is still solvable since the number
of excitations above the fully polarized ground state become
very small. Let us focus on
$h\sim h_{c1}$, the solution for $h\sim h_{c2}$ can be deduced by
symmetry.

Another useful way of viewing this problem, specially useful when we
deal with the low density limit, is given by using the boson
representation of spins, instead of the standard Jordan-Wigner fermionic
one (the fermions have to carry a string).
The spins can be represented by hard core bosons. The presence of a
boson denotes a spin up state or in the original ladder a triplet on the
rung. The hard core constraint ensures that one has only two states
(full or empty) on each site to get a faithful representation of the
spin 1/2. The problem thus reduces to a problem of hard core bosons
with in chain interactions (due to the $\SS^z \SS^z$ term). The
interchain coupling Hamiltonian is thus just the kinetic energy
interchain hopping term of these bosons
\begin{equation} \label{eq:}
\frac{J'}4 \sum_{i,\langle \alpha,\beta \rangle}b^\dagger_{i,\alpha}
 b_{i,\beta}
\end{equation}

Since the bosons are very diluted it is essentially exact to
neglect the interactions between them but for the hard core constraint,
as indicated by the fact that the Luttinger liquid parameter for a
single chain goes to $K=1$ close to $h_{c1}$. One has thus to solve the
problem of a three dimensional gas of hard core bosons with the simple
kinetic energy
\begin{eqnarray} \label{eq:bosonlow}
H_{\text{boson}} & = & \frac{J_\parallel}2 \sum_{i,\alpha}
(b^\dagger_{i,\alpha} b_{i+1,\alpha} + \text{h.c.}) \\
& + & \frac{J_\perp}2 \sum_{i,\langle \alpha,\beta \rangle}
(b^\dagger_{i,\alpha} b_{i,\beta} + \text{h.c.}) \nonumber
\end{eqnarray}
To go back to the standard negative hopping one must
perform along the chain the gauge transformation
\begin{equation} \label{eq:gauge}
c_i \to (-1)^i c_i
\end{equation}
Since the density is low (\ref{eq:bosonlow}) can be reduced to
the continuum limit of bosons with the kinetic energy
\begin{equation}
E(k,k_\perp) = \frac{k_\parallel^2}{2 m} + \frac{k^2}{2 M}
\end{equation}
The ordering transition thus reduces to the well known problem of the
Bose condensation transition for the diluted boson gas
\cite{popov_functional_book}.
The three dimensional ordered phase corresponds to the superfluid one,
whereas the high temperature phase is the normal fluid of bosons.
Most physical quantities relevant for the ladder problem can
immediately be borrowed from the vast knowledge existing for the
diluted boson gas. We will not dwell on all quantities that
can be computed but simply give here a few examples.

The critical temperature is known in the limit of low density and
is given by the equation \cite{popov_functional_book}
\begin{equation} \label{eq:lamdef}
\frac{\Lambda}{t_0} = \frac{\bigh}{t_0} - \frac{2}{(4\pi)^{3/2}} \zeta(3/2)
T^{3/2} = 0
\end{equation}
where $t_0$ is a simple number (scattering matrix for an infinite hard core
potential).
This leads to a critical temperature varying as
\begin{equation} \label{eq:tchc1}
T_c \propto (h - h_{c1})^{2/3}
\end{equation}
and gives back immediately the mean field critical exponents.
Of interest is also the total density $\rho$, i.e. the magnetization
of the ladder. It is given by
\begin{eqnarray}
\rho & = & \frac{\bigh}{t_0} - \frac{1}{(4\pi)^{3/2}} \zeta(3/2)
T^{3/2} \qquad,\qquad T < T_c \\
\rho & = & \frac{1}{(4\pi)^{3/2}} \zeta(3/2)
T^{3/2} \qquad,\qquad T > T_c \label{eq:totdens}
\end{eqnarray}
Thus two effects occurs. First the magnetization is non monotonous in
temperature and increases by a factor of two between $T=T_c$ and
$T=0$. Second,  at criticality $h=h_{c1}$ the
magnetization grows as $T^{3/2}$ at very low temperatures
(which can be readily seen by
computing $\int d^dq n_B(\epsilon(q))$. This is a different temperature
dependence than for \emph{independent} ladders. In that case it
would be given by the excitations of the one-dimensional theory, which
are fermionic in nature and have a one dimensional density of state,
leading to
\begin{equation} \label{eq:indep}
m \propto \int_0^{\epsilon_{\text{max}}} d\epsilon N_{1d}(\epsilon)
n_F(\epsilon) \propto T^{1/2}
\end{equation}
Looking at the temperature dependence (for very low temperatures)
of the the magnetization at
criticality should thus provide information on the interladder
coupling. Of course at higher temperatures one always recovers the
independent ladders behavior (\ref{eq:indep}).

Another important quantity is the NMR relaxation rate. The $\langle S^+ S^-
\rangle$ correlation function is here simply given by the bosonic single
particle Green's function
\begin{equation}
\langle S^+(r,\tau) S^-(0,0)  \rangle = (-1)^r
\langle c(r,\tau) c^\dagger(0,0)  \rangle
\end{equation}
The $(-1)^r$ factor coming from the transformation (\ref{eq:gauge})
implies that the low energy (massless) part of the spectrum
which for the spins is around $q \sim \pi$
(see Figure~\ref{fig:masslessmassive}))
is given by the small $\omega$ small $q$ Green's function for the bosons.
The single particle Green's function can also be computed in the limit
of low density \cite{popov_functional_book} and is given in the
condensed low temperature phase by
\begin{equation} \label{eq:greencond}
G(\omega_n,q) = \frac{i \omega_n + k^2 + \Lambda}{\omega_n^2 + k^4 +
                      2 \Lambda k^2}
\end{equation}
Using the standard formula for the NMR relaxation rate (\ref{eq:nmrbooks})
and (\ref{eq:greencond}) one obtains
\begin{equation}
\frac1{T_1} \propto \frac{T}{\sqrt\Lambda}
\end{equation}
where $\Lambda$ is given by (\ref{eq:lamdef})
giving thus a relaxation rate proportional to $T$ at low temperature
and diverging close to $T_c$ as $1/(T_c - T)^{1/2}$.

The above results could apply to the three dimensional phase of
Cu$_2$(C$_5$H$_{12}$N$_2$)$_2$Cl$_4$, which is a strong coupling
ladder \cite{chaboussant_ladder_strongcoupling}
where this theory is directly applicable. Whether or not the
transition experimentally observed is due to the mechanism presented
here is still an open question and other mechanisms of instability have been
proposed \cite{nagaosa_lattice_ladder,calemczuk_heat_ladder}.
Various experiments can be performed to elucidate this point.
First since the 3D transition described here is simply an ordering in the
direction of the spins in the XY plane its impact on the global
global magnetization is very weak, as seems to be the case
experimentally. Note however that it does change the temperature
dependence of the magnetization at criticality and below $T_c$.
Other interesting experiments
could be a fit of the $h-T_c$ relation (\ref{eq:tchc1})
close to $h_{c1}$ and more
generally the camel-like shape of the phase diagram.
Local probes like NMR
or neutrons should be perfectly suited to study this transition.
NMR could provide a way to map the phase boundary
(by looking at the divergence of $1/T_1 \sim 1/|T_c-T|^{1/2}$).
The $1/T_1 \sim T$ law in the low temperature
phase could also provide conclusive evidence.
Finally since we know that the
transition is in the normal-superfluid transition universality class,
one could also try to compare the thermodynamic singularities.

\section{Conclusion} \label{sec:conclusion}

We have examined in this work the properties of ladders under
magnetic field, and focussed on the gapless phase occuring between
$h_{c1} < h < h_{c2}$. For a single ladder we computed
quantitatively the correlation functions as a functions of the
magnetic field. The correlation functions in the ladders have
an identical structure both for the weak coupling ladder $J_\perp \ll
J_\parallel$ and the strong coupling one. As in weak coupling an
interesting feature in the spin correlation function of the ladder
compared to a single chain is the appearance for $h \sim h_{c1}$
of a low energy mode
\emph{only} close to $q\sim 0$ for the $\langle S_z S_z \rangle$
correlation function and close to $q\sim \pi$ for
the $\langle S_+ S_- \rangle$ one. A single chain would have had both
$q\sim 0$ and $q\sim \pi$ mode at low energy. This prediction should
be testable in neutrons experiments.
Close to the lower critical field $h_{c1}$ where the gap
is destroyed we recovered the universal exponent
\cite{chitra_spinchains_field} for the
divergence of the NMR relaxation rate
$1/T_1 \sim 1/T^{1/2}$.

These results served as a basis to analyze the nature of the
phase transition in a system of three dimensionally coupled ladders.
This problem falls into the category of Bose condensation of hard core
bosons, which allows to obtain many properties of the phase diagram
and the ordered phase.  The variation of $T_c$ with the field
has a local \emph{minimum} at $h=(h_{c1}+h_{c2})/2$ leading to an
unusual camel-like shape for the phase diagram.
Close to $h_{c1}$ and $h_{c2}$ the
transition is similar to the one of a diluted Bose gas with
$T_c \sim (h - h_{c1})^{2/3}$. The temperature dependence of the $S_s$
magnetization goes at criticality $h=h_{c1}$ from a $T^{1/2}$ behavior
for independent ladders to a $T^{3/2}$ (when interladder coupling
intervenes) at low temperatures.
The NMR rate diverges close to the
transition as $1/T_1 \sim |T-T_c|^{-1/2}$ and behaves as $1/T_1 \sim
T$ at low temperatures. These quite distinct features
could be used to check whether this transition is the one occuring in
the experimental system Cu$_2$(C$_5$H$_{12}$N$_2$)$_2$Cl$_4$.

\acknowledgements

We thank L. P. Levy for many inspiring discussions.
A. T. is grateful to University of Paris-Sud, where part of this work was
completed, for support and hospitality. T.G. thanks the I.T.P.
(Santa Barbara), where the final stage of this work was completed,
for support and hospitality. This research was supported in part by the
National Science Foundation under grant PHY94-07194.

\appendix

\section{Weak vs strong coupling ladder} \label{ap:weaktostrong}

We show in this appendix the connection between the correlation
functions for the weak coupling ladder and the strong
coupling one. The use of the simple luttinger liquid expressions due to
Haldane \cite{haldane_bosons} allows for a more transparent derivation
than the one given in Ref.~\onlinecite{furusaki_correlations_ladder}.

For weak $J_\perp$ one introduces two boson fields (one for each leg)
and it is more
convenient to use the symmetric and antisymmetric combinations.
\begin{equation}
\phi_{1,2} = \frac{\phi_s \pm \phi_a}{\sqrt2}
\end{equation}
With the usual representations of the spin operators
in terms of the $\phi$ and $\theta$. Since the magnetic field
couples only to the symmetric field, it can only remove this gap
and the antisymmetric field remains \emph{massive} even above
$h_{c1}$. To write the correlation functions for the spins
in terms of the bosonic fields one use the standard representation
of spin in terms of fermion operators and then the expression
of this fermions in terms of the  bosonic ones for a Luttinger liquid.
Let us start with the
$S_z$ operator, connected with the density of associated fermions.

We use the decomposition of the density or fermion
operator in a Luttigner Liquid.
\begin{equation} \label{eq:higher}
\rho = \rho_0 - \frac1\pi \nabla_x\phi + \sum_n e^{i n (Q x + \phi)}
\end{equation}
which contains all harmonics $2 k_F$, $4 k_F$ etc. of the fermion
density. Traditionally one only retains the lowest (most singular)
harmonic, which leads to the standard
expression (\ref{eq:bosonizedspin}). However here, since here
the $2k_F$ component is massive due to the presence of
the antisymmetric mode, it makes it necessary to retain
the next harmonic.
This leads to the spin operators (e.g. for spins on chain $1$)
\begin{equation}
S_z  =  \nabla \phi_s  + e^{i Q x + \sqrt2(\phi_s + \phi_a))}
          + e^{i 2 Q x + 2 \sqrt2(\phi_s + \phi_a)} + \cdots
\end{equation}
Since the field $\phi_a$ remains massive even above $h_{c1}$,
all correlation functions containing it decay exponentially, and can
be neglected at large distance (or time). Thus no $Q$ component
appears in the correlation function for the ladder
\cite{chitra_spinchains_field}, in marked contrast to the frustrated
or dimerized single chain. On the other hand the $2Q$ term contains
$2\sqrt2 \phi_a$. Although this term is superficially massive, it can
be combined with a $\cos(2\sqrt2 \phi_a)$ term existing in the
Hamiltonian for the ladder, giving rise to the operator
\begin{equation}
C e^{i (2 Q x + 2 \sqrt2(\phi_s))}
\end{equation}
where $c$ is a mere constant. This operator containing only the
symmetric field is massless. The long wavelength decay of the
correlation function in the weak coupling ladder is thus given by
\begin{equation} \label{eq:asweakcoupl}
\langle S_z(r) S_z(0) \rangle = \frac1{r^2} + C^2 \cos(2 Q)
                     \left(\frac1r\right)^{4K}
\end{equation}
For the weak coupling ladder since $K \ge 1/2$ the $\cos(2Q)$ term is
subdominant and can be safely dropped. To make the connection with
the case of strong coupling, where one can have $K < 1/2$, it must be
kept \cite{furusaki_correlations_ladder}.
Since $Q=\pi(1-2m)$ it is easy to see that (\ref{eq:asweakcoupl})
has exactly the same form than the expression derived for
the strongly coupled ladder (\ref{eq:corladderstrong}),
showing that the two limits
are smoothly connected.

Similar results can be obtained for the higher harmonics $2 n Q$.
For the transverse magnetization correlation one get in a similar way
for the weak coupling ladder
\begin{equation}
S^+ = e^{i\theta}\left[(-1)^i + \cos(2\phi) + (-1)^i \cos(4\phi)\right]
\end{equation}
where the $\cos(4\phi)$ term comes again from the higher harmonics.
As for the $S_z$ component, the $\cos(2\phi)$ remains massive due to the
gap in the $\phi_a$ field, whereas the $\cos(4\phi)$  can again be
combined with terms in the Hamiltonian to give a massless term.
The final result is
\end{multicols}
\begin{equation}
\langle S^+(r) S^-(0) \rangle = (-1)^r \left(\frac1{r}\right)^{1/4K}
                    + \cos(\pi(1+2m)r) \left(\frac1r\right)^{1/4K+4K}
\end{equation}
\begin{multicols}{2}
Thus the expression (\ref{eq:corladderstrong}) for the
strongly coupled ladder  is again similar to this one.

Thus weak and strong coupling ladders are smoothly connected. The
crucial reason is that the gap in the antisymmetric degrees of freedom
which exists already in the weak coupling ladder is equivalent to the
neglect of the two excited states of the triplet performed for the
strong coupling. In a system without such an antisymmetric gap (such
as a dimerized chain) this smooth continuity would not hold and the
weak and strong coupling correlation functions would be radically different.


\end{multicols}
\end{document}